\documentclass[12pt, a4paper]{article}
\usepackage{jheppub}
\usepackage{bm}
\usepackage{bbm}
\usepackage{multirow}
\usepackage{amsmath}
\usepackage{amssymb}
\usepackage{amsfonts}
\usepackage{mathrsfs}
\usepackage{graphicx}
\usepackage{grffile}
\usepackage[11pt]{moresize}
\usepackage[normalem]{ulem}
\usepackage[dvipsnames]{xcolor}
\usepackage[]{hyperref}
\usepackage{orcidlink}
\usepackage{relsize}
\usepackage{placeins}
\usepackage{makecell}
\usepackage{lineno}
\usepackage{appendix}
\usepackage{mathtools}
\usepackage{subcaption}
\usepackage{overpic}
\usepackage{tablefootnote}
\usepackage{halloweenmath}
\usepackage[export]{adjustbox}
\usepackage{placeins}
\usepackage[utf8]{inputenc}
\usepackage{wasysym}

\hypersetup
{
 bookmarks  = true,           
 unicode    = false,          
 pdfcreator = {RevTeX},       
 colorlinks = true,           
 linkcolor  = red,            
 citecolor  = blue,           
 filecolor  = black,          
 urlcolor   = blue,           
}

\newcommand{\capdef}{}
\newcommand{\mycaption}[2][\capdef]{\renewcommand{\capdef}{#2}
\caption[#1]{{\footnotesize #2}}}
\makeatletter
\renewcommand{\fnum@table}{\textbf{\tablename~\thetable}}
\renewcommand{\fnum@figure}{\textbf{\figurename~\thefigure}}
\makeatother

\newenvironment{conditions*}
{\par\vspace{\abovedisplayskip}\noindent
  \tabularx{\columnwidth}{>{$}l<{$} @{${}={}$} >{\raggedright\arraybackslash}X}}
{\endtabularx\par\vspace{\belowdisplayskip}}

\date{\today}

\title{Probing long-range $L_e-L_\mu$ forces with supernova neutronization burst neutrinos}

\author[a] {\orcidlink{0000-0001-6639-0951}Amol Dighe,}
\author[b] {\orcidlink{0000-0001-6719-7723}Sadashiv Sahoo,}
\author[b] {\orcidlink{0000-0001-7948-4332}Manibrata Sen}

\affiliation[a]{Tata Institute of Fundamental Research,\newline Homi Bhabha Road, Colaba, Mumbai 400005, India}
\affiliation[b]{Department of Physics, Indian Institute of Technology Bombay,\newline Powai, Mumbai 400076, India}

\emailAdd{amol@theory.tifr.res.in}
\emailAdd{sadashiv@iitb.ac.in}
\emailAdd{manibrata@iitb.ac.in}

\abstract{Ultralight gauge bosons associated with flavour-dependent leptonic symmetries generate long-range potentials that can modify neutrino flavour evolution over astrophysical distances. We investigate the sensitivity of neutronization-burst neutrinos from core-collapse supernovae for such interactions in the anomaly-free $U(1)'_{L_e-L_\mu}$ framework. Incorporating the long-range potential into supernova neutrino oscillations, we simulate the corresponding signal in the Deep Underground Neutrino Experiment (DUNE) using a realistic detector response of its 40 kt Liquid Argon Time Projection Chamber. We show that in the range where the long-range potential dominates over or is comparable to the vacuum oscillation term, the electron-neutrino survival probability can be significantly modified. This would produce observable distortions in the time and energy distributions of the neutronization burst neutrino spectra. Our results demonstrate that future observations of galactic supernova neutrinos, particularly from a nearby event such as Betelgeuse, can provide a sensitive and complementary probe of flavour-dependent long-range leptonic interactions.}


\begin{document}
\maketitle

\section{Introduction and Motivation}
\label{sec:intro}
The discovery of neutrino oscillations has established that neutrinos possess finite masses and undergo flavour mixing, providing the first experimentally confirmed evidence of physics beyond the Standard Model (BSM)~\cite{PDG:2024cfk}. Over the next few decades, a large number of neutrino experiments have measured the parameters governing flavour oscillations with excellent precision, thereby transforming neutrinos into one of the most sensitive probes of new physics. Despite this progress, the origin of neutrino masses and flavour structure remains unknown, necessitating extensions of the Standard Model (SM). 

Neutrinos produced in astrophysical environments play a crucial role in this program. In particular, supernova (SN) neutrinos travel through highly dense matter, making them sensitive to subtle modifications of the standard flavour evolution. These neutrino sources provide powerful laboratories for testing neutrino properties and searching for new interactions that may be inaccessible in terrestrial experiments. The only direct observation of these neutrinos happened in 1987, when three neutrino observatories, namely, Kamiokande-II~\cite{Kamiokande-II:1987idp}, IMB~\cite{Bionta:1987qt}, and Baksan (BST)~\cite{Alekseev:1988gp} independently observed a burst of neutrinos associated with SN1987A in the Large Magellanic Cloud, which is at a distance of nearly 50 kpc from the solar system. Although only a few dozen events were observed, this detection confirmed the basic theoretical picture of stellar core collapse, and established the foundations of MeV neutrino astronomy. Modern neutrino detectors are expected to record orders of magnitude more events from the next galactic SN, opening unprecedented opportunities to study both SN physics and fundamental neutrino properties.

Among the different phases of neutrino emission, the neutronization burst phase, characterized by the first few tens of milliseconds after core bounce, is particularly valuable as a probe of new physics. This phase is dominated by electron neutrinos, produced due to electron capture on free protons. Due to the large hierarchy among the neutrino flavours, the evolution during this phase is completely dominated by the adiabatic Mikheyev-Smirnov-Wolfenstein (MSW) conversion~\cite{Wolfenstein:1977ue, Mikheev:1986gs, Mikheev:1986wj}. Consequently, the emitted neutrino signal is expected to predominantly emerge as the $\nu_3$ mass eigenstate for normal mass-ordering (NMO), and as the $\nu_2$ mass eigenstate for inverted mass-ordering (IMO). This remarkable feature provides an exceptionally clean laboratory for testing fundamental neutrino properties, since deviation from this picture could reveal the presence of non-standard interactions (NSIs) of neutrinos.

A large class of BSM models predict the existence of flavour-dependent long-range interactions (LRIs) among leptons. Of particular interest are long-range leptonic interactions mediated by ultralight scalar bosons~\cite{Chattopadhyay:2026wil, Choi:2019zxy,Ge:2020xkm,Ge:2020ffj,Choi:2020ydp,Smirnov:2021zgn,Chun:2021ief,Sen:2023uga,Ge:2024ftz,Perez-Gonzalez:2025qjh,Pompa:2025lbf,Chattopadhyay:2025ccy, Davoudiasl:2023uiq,Lopes:2023vxn,Martinez-Mirave:2024dmw,Goertz:2024gzw}, and vector bosons associated with additional $U(1)'$ gauge symmetries, such as $(L_e-L_\mu)$, $(L_e-L_\tau)$ and $(L_\mu-L_\tau)$, which can modify neutrino oscillation phenomenology through the introduction of flavour-dependent potentials. Their consequences have been explored using solar~\cite{Grifols:2003gy,Bandyopadhyay:2006uh,Gonzalez-Garcia:2006vic}, atmospheric~\cite{Joshipura:2003jh,Garg:2026gwx}, long-baseline~\cite{Heeck:2010pg,Heeck:2018nzc,Jana:2026pwx, Agarwalla:2024ylc}, and astrophysical neutrino observations~\cite{Bustamante:2018mzu,Agarwalla:2023sng}, as well as through global analyses of oscillation data~\cite{Davoudiasl:2011sz,Farzan:2016wym,Wise:2018rnb,Dror:2020fbh,Coloma:2020gfv,Alonso-Alvarez:2023tii}. Additional constraints arise from gravitational and precision tests~\cite{Schlamminger:2007ht,Adelberger:2009zz,Salumbides:2013dua,Baryakhtar:2017ngi,KumarPoddar:2019ceq,KumarPoddar:2020kdz}. While no experimental indication of such interactions has yet been observed, the surviving parameter space continues to motivate searches in environments where even feeble long-range forces may leave observable signatures.

SN neutrinos during the neutronization burst provide a qualitatively distinct probe of such interactions. While adiabatic propagation of these neutrinos implies that any smooth potential added to the MSW potential inside the star will not modify the neutrino mass composition arriving at the Earth, the neutrino flavour composition arriving at the Earth may be quite different when this additional potential is significant near the Earth. In particular, if ultralight particles mediate the potential and the coupling is large, the effects of electrons inside the SN may extend all the way from the SN to the Earth, and hence alter the propagation picture completely. In such a scenario, neutrinos may never really propagate in ``vacuum''; rather a new neutral-current NSI potential induced due to the long-range interactions persists all along. In this limit, neutrinos would never really deviate much from the flavour eigenstates and hence would arrive at the Earth almost as flavour eigenstates. Even if the effects due to the electrons inside the SN become insignificant due to the large distance, the effects of the electrons inside the Sun - or the combined effects of electrons from the whole galaxy and beyond - may be appreciable near the Earth. Thus, LRIs can alter the electron neutrino survival probability, and hence the observable event distribution in neutrino detectors, providing a direct signature of new physics. 

Motivated by this prospect, in this article, we investigate flavour-dependent LRIs using the neutronization-burst neutrino signal from a prospective core-collapse SN. We consider the minimal anomaly-free gauge extension of the SM based on the $L_e-L_\mu$ symmetry, evaluate the impact of the induced potential on SN neutrino flavour conversion, and determine the corresponding (prospective) signatures in the Deep Underground Neutrino Experiment (DUNE).

As a benchmark source, we consider Betelgeuse, a nearby red supergiant located at a distance of approximately $168~{\rm pc}$ and widely regarded as one of the most promising candidates for a future galactic core-collapse supernova. Owing to its proximity, a Betelgeuse explosion would produce an enormous neutrino signal, enabling precision studies of flavour conversion effects that would otherwise be inaccessible. Our analysis demonstrates the potential of neutronization-burst observations as precision probes of ultraweak flavour-dependent long-range interactions and exemplifies the discovery and exclusion prospects of future neutrino observatories in the MeV-energy neutrino sector.

The remainder of this article is organized as follows. Section~\ref{sec:flri} discusses the formalism for long-range interactions mediated by a new symmetry. Section~\ref{sec:form} focuses on neutrinos from a SN, and how they are impacted by the long-range interactions. Section~\ref{sec:DUNEinto} presents an analysis of the neutrino spectra in a DUNE-like detector setup, and section~\ref{sec:result} discusses the sensitivity of such an experiment to the LRIs. Finally, we conclude in section~\ref{sec:cnrmk}.

\section{Flavour-dependent long-range leptonic interactions}
\label{sec:flri}
The SM is known to conserve the total lepton number at a classical level, however, neutrino oscillations violate individual flavour lepton numbers $\left(L_e,\,L_\mu,\,L_\tau\right)$, thereby pointing to a non-trivial flavour structure. Since the minimal SM cannot accommodate massive neutrinos, numerous extensions have been proposed to elucidate the underlying origin of neutrino masses and flavour dynamics. Among the most economical and theoretically well-motivated scenarios are gauge extensions involving an additional abelian symmetry $U(1)'_X$, where the charge assignment $X$ is chosen such that the enlarged gauge group $SU(3)_{\rm c}\otimes SU(2)_{\rm L}\otimes U(1)_{\rm Y}\otimes U(1)'_X$ remains anomaly-free, gauge-invariant, and renormalizable~\cite{Foot:1990mn,He:1991qd,Foot:1994vd}.

In purely leptonic constructions, $X$ may correspond to anomaly-free combinations of lepton flavour numbers, such as $L_e-L_\mu$, $L_e-L_\tau$, or $L_\mu-L_\tau$. Gauging any of these symmetries predicts the existence of a new electrically neutral vector boson, $Z'$, which mediates flavour-dependent interactions and can induce observable effects in neutrino propagation and oscillation phenomena~\cite{Langacker:2008yv}. The phenomenology of these interactions depends on $m_{Z'}$, the mass of $Z'$. For mediator masses comparable to the electroweak scale, the interaction is effectively short-ranged and is commonly described within the framework of non-standard neutrino interactions (NSIs). On the other hand, if the mediator is ultralight, the interaction range can become macroscopic and may extend over astronomical distances. 

In this regime, neutrinos propagating through the Universe can experience flavour-dependent potentials generated by distant astrophysical sources. For the gauge symmetries depending on a combination of lepton numbers, the associated neutrino matter potential can be sourced predominantly by electrons\footnote{This happens because muons and taus are essentially absent in ordinary matter.}. The interaction can lead to a Yukawa-like potential
~\cite{Gonzalez-Garcia:2006vic, Bandyopadhyay:2006uh} outside the SN:
\begin{align}
	\mathcal{V}_e(\vec{r}) = \frac{{g'_e}^2}{4\pi} \int_{0}^{R_{\rm SN}} d^3 x^\prime\; n_e(\vec{x}^{\,\prime})\; \frac{e^{-|\vec{x}^{\,\prime} - \vec{r}|/\lambda}}{|\vec{x}^{\,\prime} - \vec{r}|} \; ,
	\label{Eq:1.2}
\end{align}
where $g'_e$ denotes the gauge coupling, $n_e(\vec{x}^{\,\prime)}$ is the electron number density, and
\begin{equation}
\lambda = \frac{1}{m_{Z'}}
\end{equation}
is the interaction range associated with the mediator mass $m_{Z'}$.
Throughout this work, we consider the LRI potential is sourced by an underlying $U(1)'_{L_e\,-\,L_\mu}$ symmetry.
For a spherically symmetric matter distribution and for $r \gg R_{\rm SN}$,
\begin{align}
	\mathcal{V}_e(r) & = \frac{{g'_e}^2}{r}\cdot e^{-r/\lambda} \cdot N_e,
	\label{Eq:1.4}
\end{align}
where $N_e$ is the total number of electrons contained within the source. This expression explicitly illustrates the interplay between the coupling strength $g'_e$ and the mediator mass $m_{Z'}$: a smaller mediator mass increases the interaction range and allows distant sources to contribute to the potential experienced by propagating neutrinos.

\begin{figure}[t]
\centering
\includegraphics[width=0.75\linewidth, trim=0 45 0 0, clip, frame]{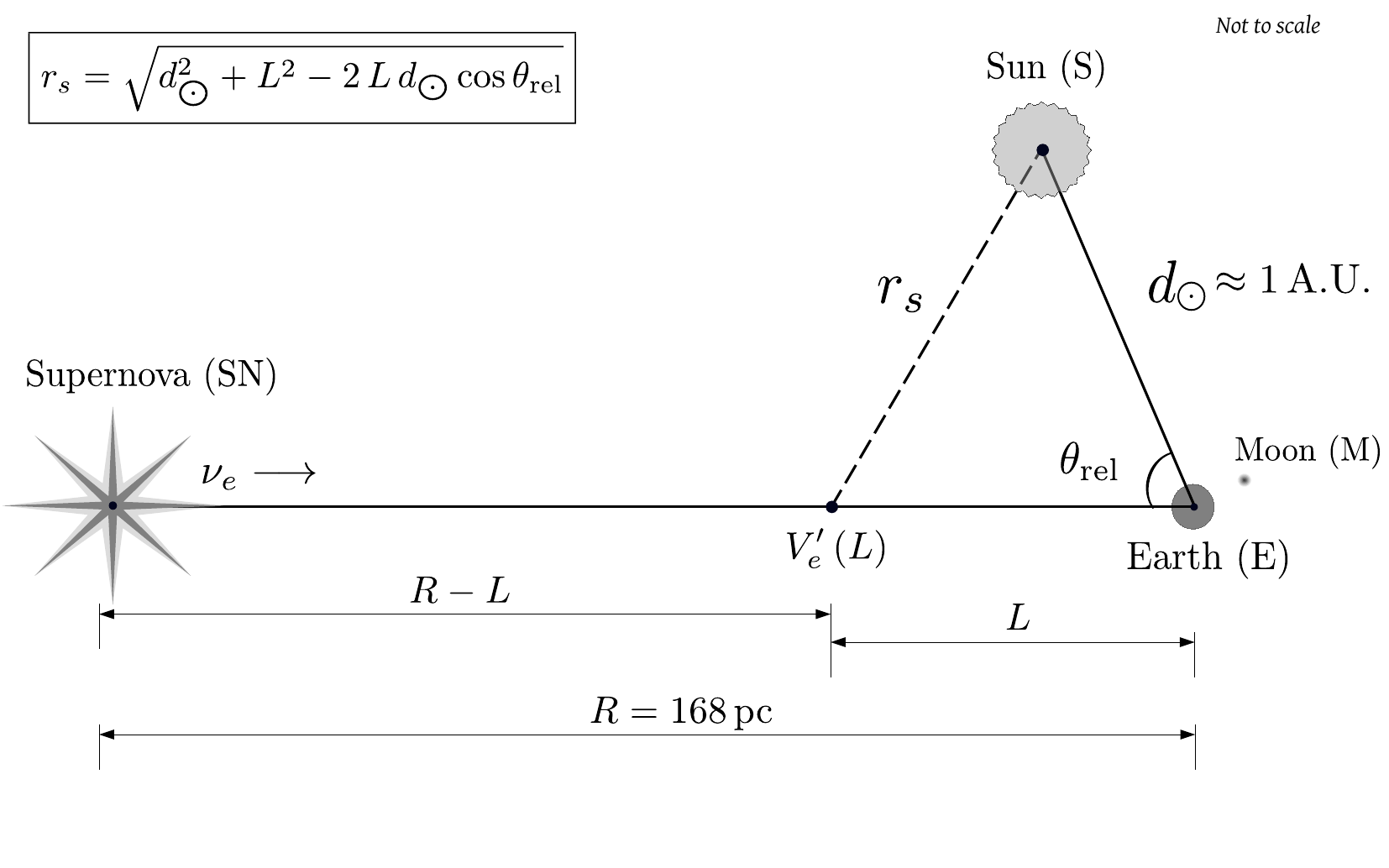}
\mycaption{A schematic illustration of the relative orientation of the Sun with respect to the Earth-SN line of sight. The Earth-SN distance is assumed to be $R=168$ pc. The quantity $V'_e$ denotes the total LRI potential at an arbitrary point located along the SN-Earth line at a distance $L$ from the center of the Earth, arising from contributions of the SN at a distance $(R-L)$ and the Sun at distance $r_s$.}
\label{fig:lsdv1}
\end{figure}

For neutrinos coming from a galactic SN, the long-range of the new force implies that contributions to the matter potential exist from electrons in the progenitor SN, the Sun, the Earth, the Moon, the Milky Way (MW) galaxy, and the extragalactic (EG) region. The total potential at the detector can therefore be expressed as 
\begin{align}
	V'_e
	&\;\approx\; \mathcal{V}_{\rm SN}\,+\,
	\mathcal{V}_{\rm Sun}\,+\,\mathcal{V}_{\rm Earth}\,+\, \mathcal{V}_{\rm Moon}\,+\, \mathcal{V}_{\rm MW}\,+\, \mathcal{V}_{\rm EG}. \label{Eq:VT}
\end{align}
The relative importance of the contributions depends on the interaction range and the geometrical configuration of the source-detector system, as shown in figure~\ref{fig:lsdv1}. 

\begin{figure}[t]
\centering
\includegraphics[width=0.75\linewidth]{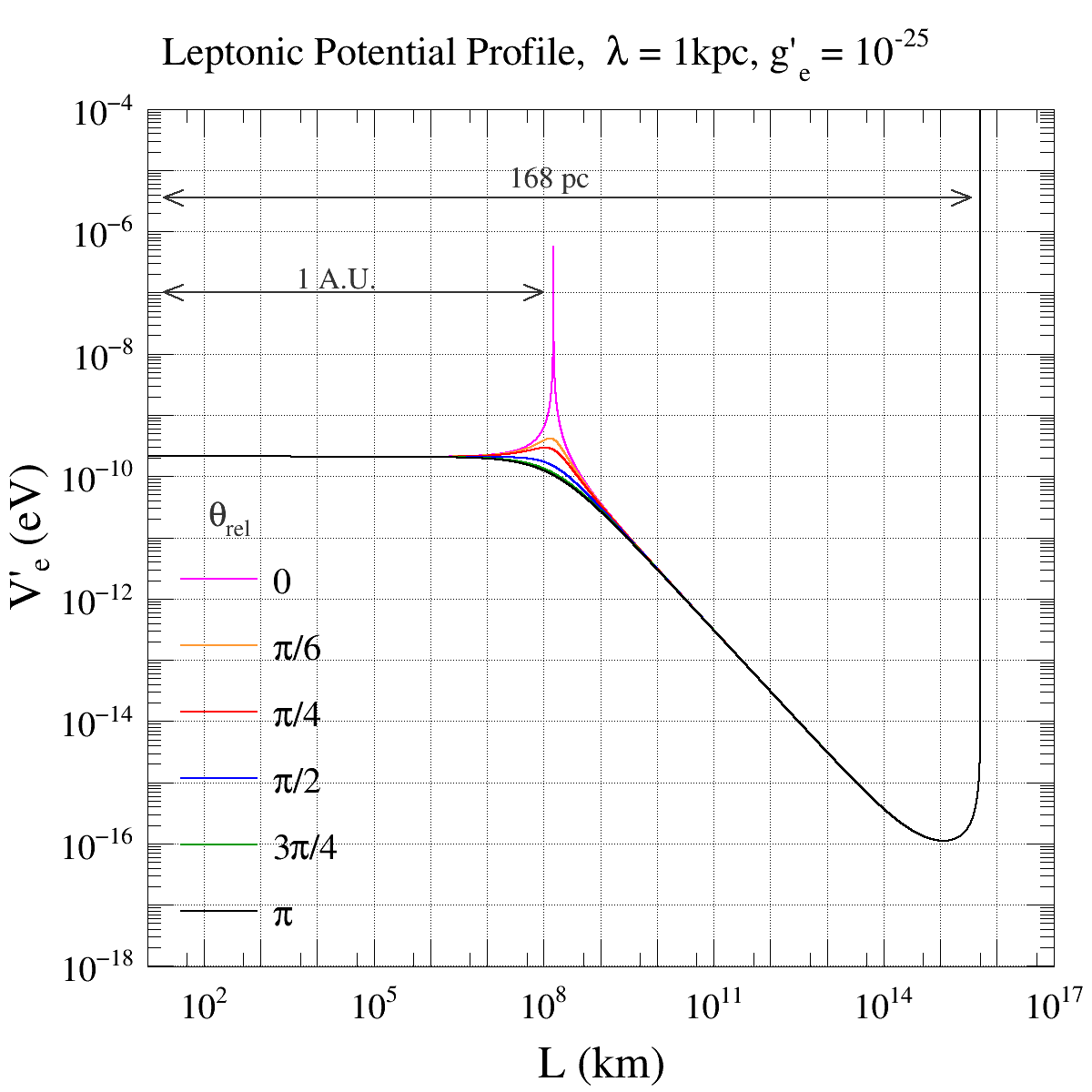}
\mycaption{The total leptonic long-range interaction potential generated by the combined contributions of the SN, the Sun, the Earth, the Moon, and the Milky Way galaxy for an effective gauge coupling $g'_e = 10^{-25}$ and an interaction range $\lambda = 1~{\rm kpc}$, as a function of distance from the Earth. The coloured curves correspond to different relative angles between the Sun and the Earth-SN line of sight $(\theta_{\rm rel})$.}
\label{fig:sdv1}
\end{figure}

Figure~\ref{fig:sdv1} illustrates the total long-range potential for a representative benchmark point with $\lambda = 1~{\rm kpc}$ and $g'_e = 10^{-25}$ for a Betelgeuse-like star located at $d=168~{\rm pc}$. The different curves correspond to various relative angles between the Earth-Sun axis and the Earth-SN line of sight\footnote{Note that Betelgeuse is located at an ecliptic latitude $\approx -16^\circ$. Consequently, the relative angle ($\theta_{\rm rel}$) varies between approximately $16^\circ$ and $164^\circ$ over an Earth annual cycle.}. When the SN lies close to the solar direction (relative angles approaching zero), the neutrino trajectory passes near the Sun, and the solar contribution becomes maximal, producing a pronounced enhancement near the Earth-Sun distance of approximately $1~{\rm AU}$. As the relative angle increases, the minimum distance between the trajectory and the Sun grows, suppressing the solar contribution. For configurations in which the SN is nearly opposite to the solar direction (relative angle approaches $\pi$), the peak becomes subdued. Thus, the long-range leptonic potential depends on the Earth-Sun-SN geometry, as seen in figure~\ref{fig:sdv1}. However, the net LRI potential observed at the Earth is almost insensitive to this relative angle ($\theta_{\rm rel}$), and for adiabatic propagation, the flavour conversion probabilities are independent of the intermediate variations in the potential.

\section{Probing LRI with supernova neutrinos}
\label{sec:form}
\subsection{Core-collapse supernova and the neutronization burst}
\label{sec:ccsn}
Massive stars end their evolution through gravitational core collapse, producing a core-collapse supernova (CCSN) and releasing nearly $99\%$ of their gravitational binding energy in the form of neutrinos~\cite{Janka:2017vlw,Mirizzi:2015eza}. During the first $\sim25$ ms after core bounce, the neutrino emission is dominated by the \textit{neutronization burst}, a robust feature of CCSN simulations~\cite{Janka:2006fh, Sen:2024fxa, Raffelt:2025wty}. This burst originates primarily from rapid electron capture on free protons, following the dissociation of heavy nuclei by the outgoing shock wave. Although thermal processes such as $e^+e^-$ annihilation and nucleon--nucleon bremsstrahlung also produce $\bar{\nu}_e$, $\nu_x$, and $\bar{\nu}_x$ ($x=\mu,\tau$), their contributions remain subdominant, rendering the neutronization burst an almost pure $\nu_e$ source. 

As the neutrinos propagate through the stellar envelope, their flavour evolution is governed by vacuum mixing, matter-induced effects, and neutrino-neutrino interactions~\cite{Duan:2010bg,Chakraborty:2016yeg,Sen:2024fxa,Raffelt:2025wty}. During the neutronization-burst epoch, the large electron density in the SN environment generates a dominant matter potential through charged-current interactions of $\nu_e$, while the collective effects are insignificant~\cite{Chakraborty:2014nma}\footnote{For a discussion on when collective oscillations can become important during the neutronization burst, see~\cite{Das:2017iuj}.}. Consequently, the neutronization burst possesses a flavour composition that is considerably cleaner and better understood than that of the later accretion and cooling phases.
The unoscillated neutrino flux emitted during the neutronization burst can be parameterized as~\cite{Keil:2002in}
\begin{equation}
	\Phi_\nu(E,t) = \frac{L_\nu(t)}{\langle E_\nu \rangle}\,\phi(E),
\end{equation}
where $L_\nu(t)$ denotes the neutrino luminosity at post-bounce time $t$ and $\langle E_\nu\rangle$ is the corresponding average neutrino energy. The spectral shape is commonly described by a pinched quasi-thermal distribution,
\begin{equation}
	\phi(E) = \frac{1}{\langle E_\nu \rangle}
	\frac{(\alpha+1)^{\alpha+1}}
	{\Gamma(\alpha+1)}
	\left(\frac{E}{\langle E_\nu \rangle}\right)^\alpha
	\exp\left[-(\alpha+1)\frac{E}{\langle E_\nu \rangle}\right],
\end{equation}
where $\alpha$ is the spectral pinching parameter and $\Gamma$ denotes the Gamma function. Both $\alpha$ and $\langle E_\nu\rangle$ evolve with time and encode the thermodynamic evolution of the exploding star~\cite{Garching}. 

An important consequence of adiabatic flavour conversion is that the neutrino produced as an electron flavour propagates as the heaviest instantaneous matter eigenstate inside the SN. After exiting the stellar envelope, this state maps onto a specific vacuum mass eigenstate. Consequently, the neutronization burst neutrino emerges from the SN as an almost pure mass eigenstate rather than as a coherent superposition of mass states. The identity of this mass eigenstate depends on the neutrino mass ordering. In the standard scenario, $\nu_e$ exits the star as $\nu_3\, (\nu_2)$ in NMO (IMO). The electron-neutrino flux arriving at Earth from a SN at a distance R is given by
\begin{align}
	f_{\nu_e}(E,t)
	&\approx
	\frac{1}{4\pi R^2}
	\bigg[
	|U_{eh}|^2\,\Phi_{\nu_e}(E,t)
	+
	\Big(1-|U_{eh}|^2\Big)
	\Phi_{\nu_{x}}(E,t)
	\bigg],
\end{align}
where $x$ denotes the non-electron flavours.
Here, $U_{eh}$ denotes the element of the leptonic mixing matrix connecting the electron flavour state to the heaviest mass eigenstate with $h=3$ ($h=2$) for NMO (IMO). This demonstrates why the neutronization burst can be an exceptionally sensitive probe of LRI. Changes to the Hamiltonian due to the LRI can lead to changes in the survival probability and consequently in the event rate measured at terrestrial detectors. We will discuss this further in the following sections.

\subsection{Long-range interactions in the \texorpdfstring{$L_e-L_\mu$}{Le-Lmu} framework}
\label{sec:form1.1}
We first consider the minimal anomaly-free gauge extension of the SM based on the $U(1)'_{L_e-L\mu}$. In this scenario, the long-range potential generated by ambient electron distributions contributes differently to electron and muon flavours, thereby modifying neutrino flavour evolution during propagation.

For an ultra-relativistic neutrino, the effective Hamiltonian in the flavour basis may be written as
\begin{align}
\mathcal{H}_{\rm eff}\,=\,&\,U\cdot\,\frac{1}{2E_\nu}\text{Diag}\Big(0,\, \Delta m^2_{21},\, \Delta m^2_{31}\Big)\cdot U^\dagger \;+\; \text{Diag}\big(V_{\rm cc}\,+\, V'_{e}, \;-V'_{e},\;0\big)\,,
\label{Eq:Ham_1.1}
\end{align}
where $U$ denotes the Pontecorvo-Maki-Nakagawa-Sakata (PMNS) mixing matrix~\cite{Pontecorvo:1957qd, Maki:1962mu, Pontecorvo:1967fh} and $E_\nu$ is the neutrino energy. 

The first term describes vacuum oscillations, while the second term includes both the charged-current matter-potential ($V_{cc}=\sqrt{2} G_F\, n^{\rm SN}_e$) prescribed by Mikheyev-Smirnov-Wolfenstein (MSW) potential~\cite{Wolfenstein:1977ue, Mikheyev:1985zog, Mikheev:1986wj} and the total LRI potential, $V'_{e}$, induced due to electrons from distant sources, is obtained from Eq.~(\ref{Eq:VT}). 
For antineutrinos, the mixing matrix $U$ becomes $U^\ast$, and the potentials flip their signs.

The key feature of Eq.~(\ref{Eq:Ham_1.1}) is that the LRI introduces an additional flavour-dependent contribution that persists even outside the SN environment. Unlike the standard MSW potential, which vanishes once the neutrino leaves the stellar envelope, the LRI contribution can remain non-negligible over astrophysical distances. Consequently, the flavour composition observed at Earth depends not only on the matter profile inside the star but also on the cumulative long-range potential generated by the surrounding astrophysical environment.

For the neutronization burst, flavour evolution is greatly simplified. Under the assumption of adiabatic evolution, the neutrino remains attached to the same instantaneous matter eigenstate of production as it propagates through the stellar envelope. Therefore, the flavour composition at production becomes largely insensitive to the precise value of the long-range potential. The observable signal is therefore governed primarily by the electron neutrino survival probability at Earth, given by
\begin{align}	P_{ee}^{\rm LRI} &\approx\; \sin^2\widetilde{\theta}^E_{13},\hspace{1.85cm} \text{NMO},
\label{Eq:Pee_2}\\
	&\approx \; \sin^2\widetilde{\theta}^E_{12}\cdot\cos^2\widetilde{\theta}^E_{13}, \hspace{0.3cm} \text{IMO},
	\label{Eq:Pee_1}
\end{align}
where the quantities
$\big(\widetilde{\theta}^E_{12}$ and $\widetilde{\theta}^E_{13}\big)$
denote the effective mixing angles evaluated at Earth in the presence of the combined MSW and LRI potentials. Thus, the problem reduces to determining the effective mixing matrix in the presence of the long-range potential at the detector location.

The effective angles can be determined following references~\cite{Bandyopadhyay:2006uh, Chatterjee:2015gta}. The Hamiltonian is diagonalized through a series of consecutive rotations $\widetilde{R}_{23}$, $\widetilde{R}_{13}$ and $\widetilde{R}_{12}$, resulting in
\begin{align}
	\sin2\widetilde{\theta}^E_{23}&\;\approx\;\left[1\,+\,	\frac{\big(\cos2\theta_{23}\cdot z_1\,+\,	\sin2\theta_{23}\cdot z_2\,+\,y'_e\big)^2}{\big(\sin2\theta_{23}\cdot z_1\,-\,\cos2\theta_{23}\cdot z_2\big)^2}\right]^{-1/2},\label{Eq:mth23}\\[2pt]
	\sin2\widetilde{\theta}^E_{13}&\;\approx\;\left[1 \,+\, \frac{\left(\Omega_3-s_{13}^2-xs_{12}^2c_{13}^2-y_{cc}-y'_e\right)^2}{\left[\left(1-xs_{12}^2\right)\sin2\theta_{13}\cos\delta'_{23}	- xc_{13}\sin2\theta_{12}\sin\delta'_{23}\right]^2}\right]^{-1/2},\label{Eq:mth13}\\[4pt]
	\sin2\widetilde{\theta}^E_{12}&\;\approx\;\left[1 \,+\,	\frac{\left(\Omega_2-\Omega_1\right)^2}{\tilde{c}_{13}^{\,2}\left[\left(1-xs_{12}^2\right)\sin2\theta_{13}\sin\delta'_{23}\,+\,x c_{13}\sin2\theta_{12}\cos\delta'_{23}\right]^2}\right]^{-1/2}.
	\label{Eq:mth12}
\end{align}
Here, $x \equiv \Delta m^2_{21}/\Delta m^2_{31}$, $z_1\,\equiv\, c_{13}^2-x c_{12}^2+x s_{13}^2 s_{12}^2$, and $z_2 \,\equiv\, xs_{13}\sin2\theta_{12}\,$.

\vspace{1em}
While the above angles specify the eigenvectors of the Hamiltonian, its eigenvalues $\Omega_{1,2,3}$ may be written in terms of the dimensionless parameters $y_{cc}\equiv 2E_\nu V^E_{cc}/\Delta m^2_{31}$ and $y'_e\equiv2E_\nu V'_{e}/\Delta m^2_{31}$ that quantify the relative importance of the standard matter and long-range interaction potentials over the vacuum terms. Here $V^E_{cc}$ stands for the MSW matter-potential at Earth.
The modified mixing parameters ($\widetilde{\theta}^E_{13}$ and $\widetilde{\theta}^E_{12}$) are encoded through $\delta'_{23} \equiv \widetilde{\theta}_{23} - \theta_{23}$. The eigenvalues $\Omega_1$, $\Omega_2$, and $\Omega_3$ are:
\begin{align}
	\Omega_3 \approx \;& \frac{1}{2}\Bigg[z_3 - y'_e\,+\, \frac{\cos2\theta_{23}\cdot z_1\,+\,\sin2\theta_{23}\cdot z_2\, +\, y'_e}{\cos2\widetilde{\theta}_{23}}\Bigg],\\[4pt]
	\Omega_2 \approx \;& \frac{1}{2}\Bigg[z_3 - y'_e\,-\, \frac{\cos2\theta_{23}\cdot z_1\,+\,\sin2\theta_{23}\cdot z_2\, +\, y'_e}{\cos2\widetilde{\theta}_{23}}\Bigg],\\[4pt]
	\Omega_1 \approx \;&\frac{1}{2}\Bigg[\Omega_3 + z_4 + y_{cc} + y'_e-\frac{\Omega_3 - z_4 - y_{cc} - y'_e}{\cos2\widetilde{\theta}_{13}}\Bigg],
\end{align}
where $z_3 \equiv c^2_{13} + x c^2_{12} + x s^2_{13}s^2_{12}$, and $z_4 \equiv s^2_{13} + xs^2_{12}c^2_{13}$.

\begin{figure}[t]
	\centering
	\includegraphics[width=0.5\linewidth]{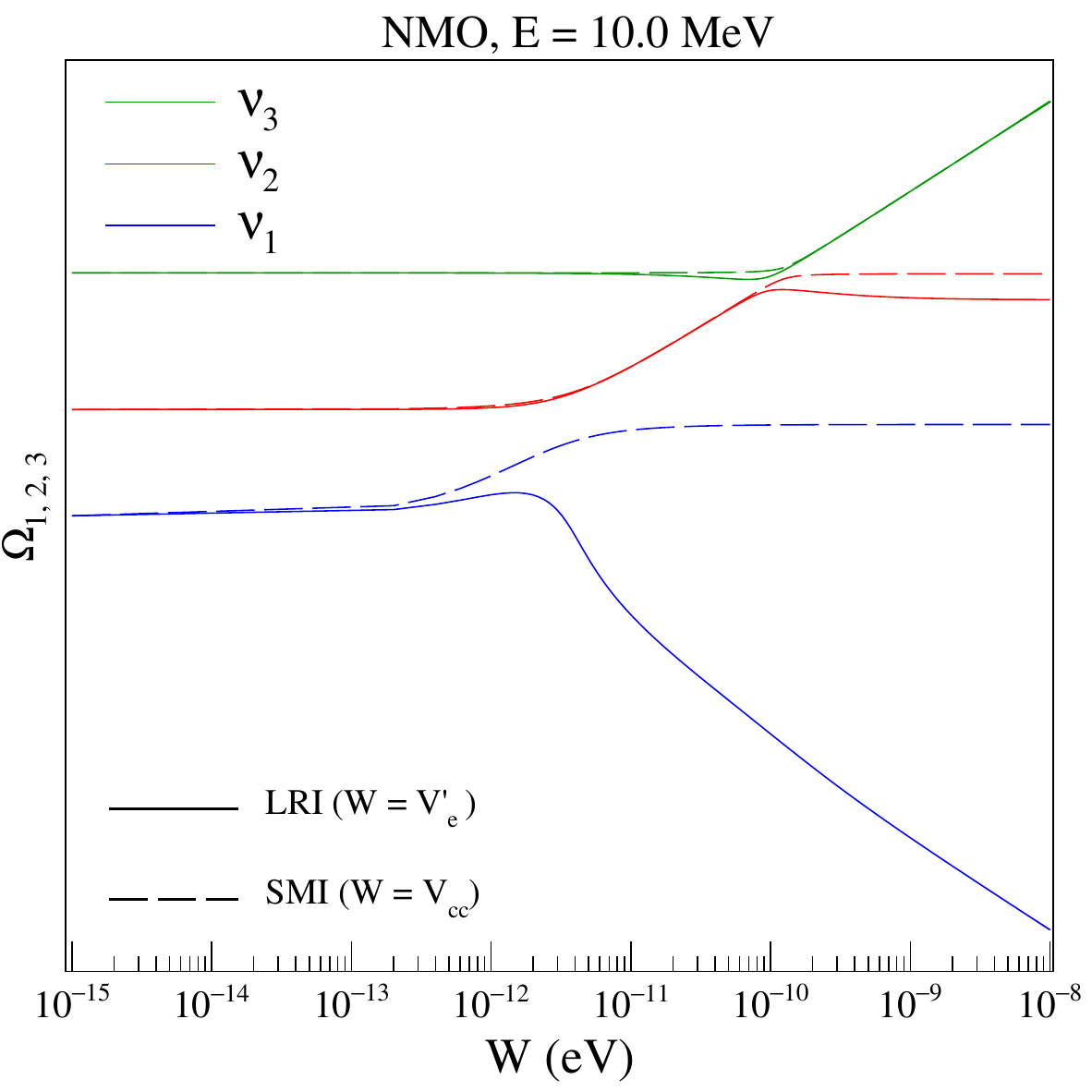}~~
	\includegraphics[width=0.5\linewidth]{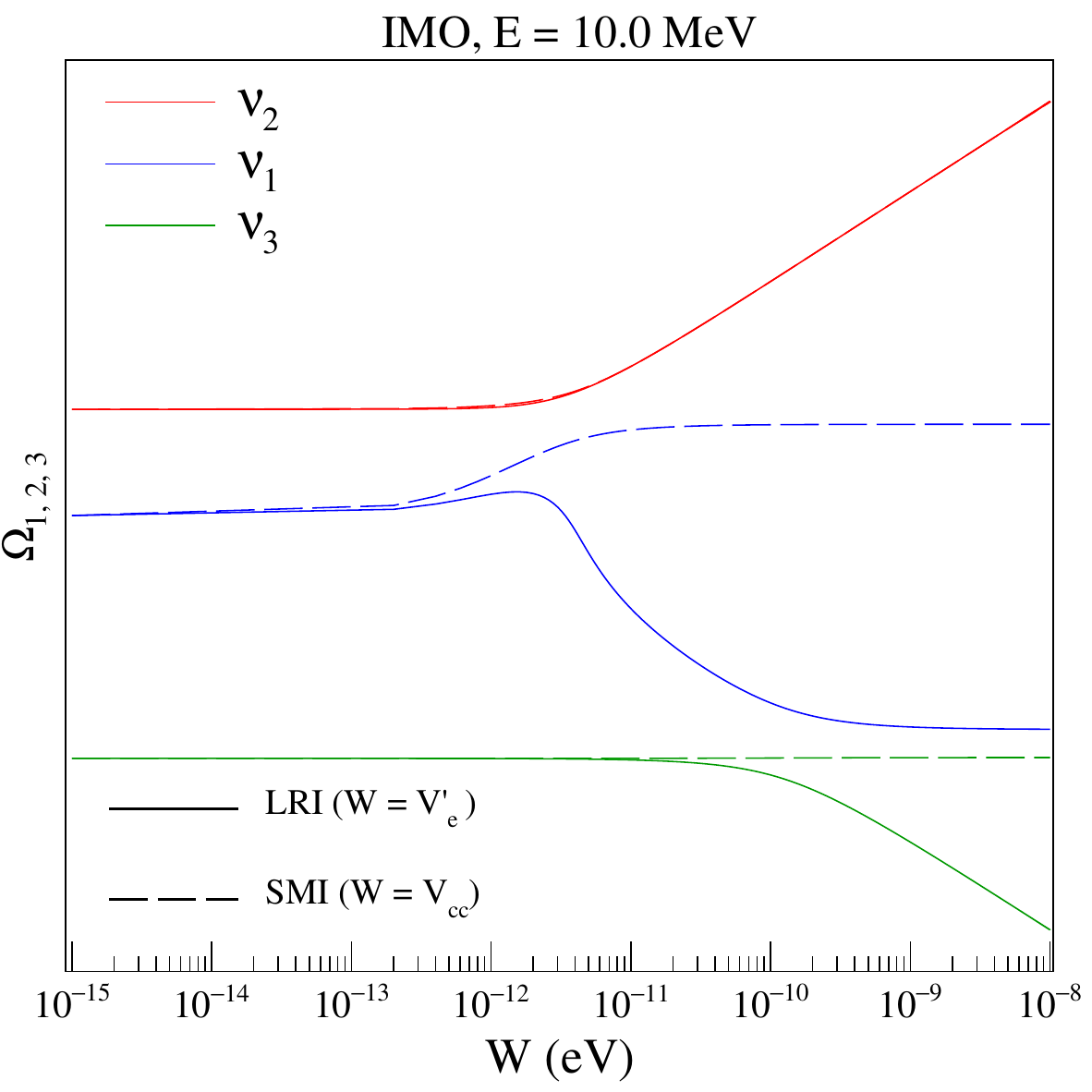}
	\mycaption{Level-crossing diagrams for the normal (left) and inverted (right) mass-ordering scenarios for a neutrino energy of $10$ MeV, as a function of the relevant potential $W$. The solid curves represent the effective mass-squared eigenvalues in the presence of an anomaly-free $L_e-L_\mu$ LRI potential only $(V_{cc}=0$, $W=V'_e$), whereas the dashed curves correspond to the standard MSW matter potential only $(V'_e=0$, $W = V_{cc}$). Note that in both the scenarios, for large values of the potential, the $\nu_e$ is produced as the heaviest matter eigenstate. However, in the LRI scenario, the asymptotic mass eigenstates at large $W$ are $\nu_\tau$ (heavier) and $\nu_\mu$ (lighter), whereas in the SMI case, the asymptotic states are linear combinations of $\nu_\mu$ and $\nu_\tau$. For animations showing the variations as a function of neutrino energies, see~\cite{animation_v}.}
	\label{fig:LCD}
\end{figure}

We depict the level-crossing diagrams for the matter eigenstates in figure\,\ref{fig:LCD} in the LRI scenario (solid lines) and compare it with the SM interactions (SMI) case (dashed lines). We find that the $\nu_e$ is produced as the heaviest matter eigenstate in both scenarios. However, due to the nature of LRI potential, the $\nu_\mu$ is produced as the lightest eigenstate, while the $\nu_\tau$ remains as the intermediate state. This is different from the SMI case~\cite{Dighe:1999bi}, where the asymptotic eigenstates at large $V_{cc}$ are specific linear combinations of $\nu_\mu$ and $\nu_\tau$. Animations depicting variations of the mixing angles and the eigenvalues as a function of the potential, and neutrino energies can be found in~\cite{animation_v}.

In the absence of LRI, the standard matter-driven three-flavour evolution implies that neutrinos can potentially encounter two resonances, called H and L, which occur when $V_{\rm H} \sim \Delta m^2_{\rm atm}/2 E_\nu \sim 10^{-10}\,{\rm eV}$ and $V_{\rm L} \sim \Delta m^2_{\rm sol}/2 E_\nu \sim 10^{-12}\,{\rm eV}$, respectively, for a typical neutrino energy $E_\nu = 10$ MeV. They encounter both these resonances in the NMO scenario, while in IMO they encounter only the L resonance. Both these resonances are expected to be adiabatic during the neutronization burst. One then gets~\cite{Dighe:1999bi}
\begin{equation}
P_{ee}^{\rm SMI}\approx|U_{e3}|^2\approx0.022,\qquad {\rm NMO},
\end{equation}
and
\begin{equation}
P_{ee}^{\rm SMI}\approx|U_{e2}|^2\approx0.30,\quad\qquad {\rm IMO}.
\end{equation}
The presence of a non-zero LRI can modify the details of the resonance structure inside the SN. However, as long as the resonances are adiabatic, the neutronization burst neutrinos exiting the star are predominantly $\nu_3$ ($\nu_2$) for NMO (IMO), just like in the case of SMI. However, on the way to the Earth they start experiencing the LRI potential due to the Sun, which keeps on increasing as depicted in figure~\ref{fig:sdv1}. This changes the effective mixing angles. The net electron-neutrino survival probability is decided by the effective mixing angles at Earth, as given by eqs.~(\ref{Eq:Pee_2}) and (\ref{Eq:Pee_1}).

In NMO, the relevant effective mixing angle is $\widetilde{\theta}_{13}^{E}$. If the LRI potential at the Earth is small, $V'_e \ll V_{\rm H}$, the LRI effect on this angle is not substantial, and hence $P_{ee}^{\rm LRI}\approx P_{ee}^{\rm SMI} \approx 0.022$. However, for $V \gtrsim V_{\rm H}$, the mixing angle can be driven to as high as $90^\circ$. Thus, $P_{ee}^{\rm LRI} \approx \sin^2 \widetilde{\theta}_{13}^{E}$ can vary continuously from its standard value of approximately $0.022$ up to values approaching unity.

In IMO, the response of the survival probability is governed by the simultaneous evolution of two effective mixing angles, $\widetilde{\theta}_{12}^{E}$ and $\widetilde{\theta}_{13}^{E}$. The LRI effects on these angles is not substantial for $V'_e \ll V_{\rm L}$, and $P_{ee}^{\rm LRI}\approx P_{ee}^{\rm SMI} \approx 0.30$. However, for $V_{\rm L} \lesssim V'_e $, the system can be close to $\widetilde{\theta}_{12}^{E} \approx 90^\circ$ and $\widetilde{\theta}_{13}^{E}\approx \theta_{13}$. The value of the survival probability can then approach $\cos^2\theta_{13}\approx 0.98$. Thus, in the IMO scenario, $P_{ee}^{\rm LRI} \approx \sin^2 \widetilde{\theta}_{12}^{E} \cos^2 \widetilde{\theta}_{13}^{E}$ can vary continuously from its standard value of approximately $0.30$ up to values approaching unity.
\begin{figure}[t]
	\centering
	\includegraphics[width=0.49\linewidth]{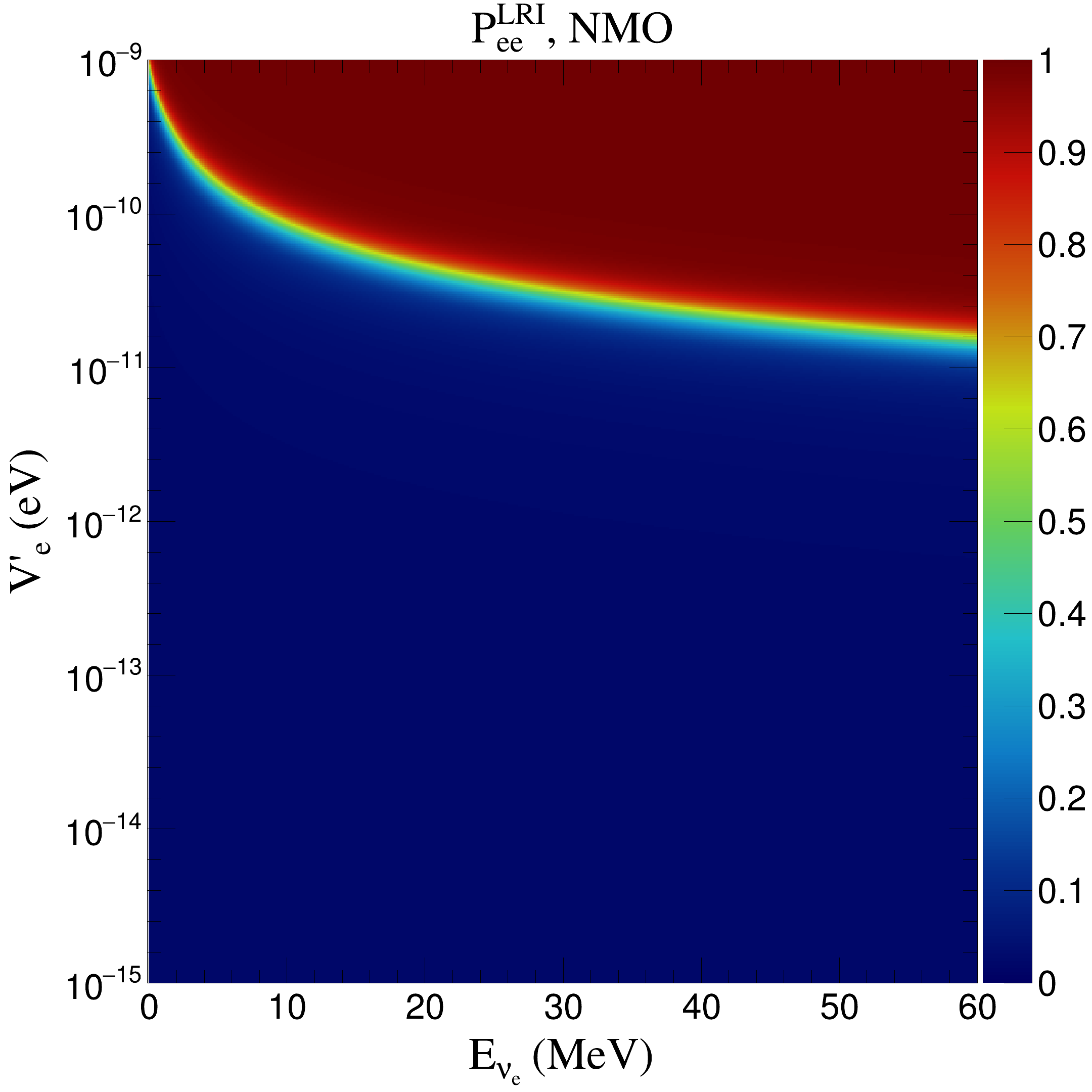}~~
	\includegraphics[width=0.49\linewidth]{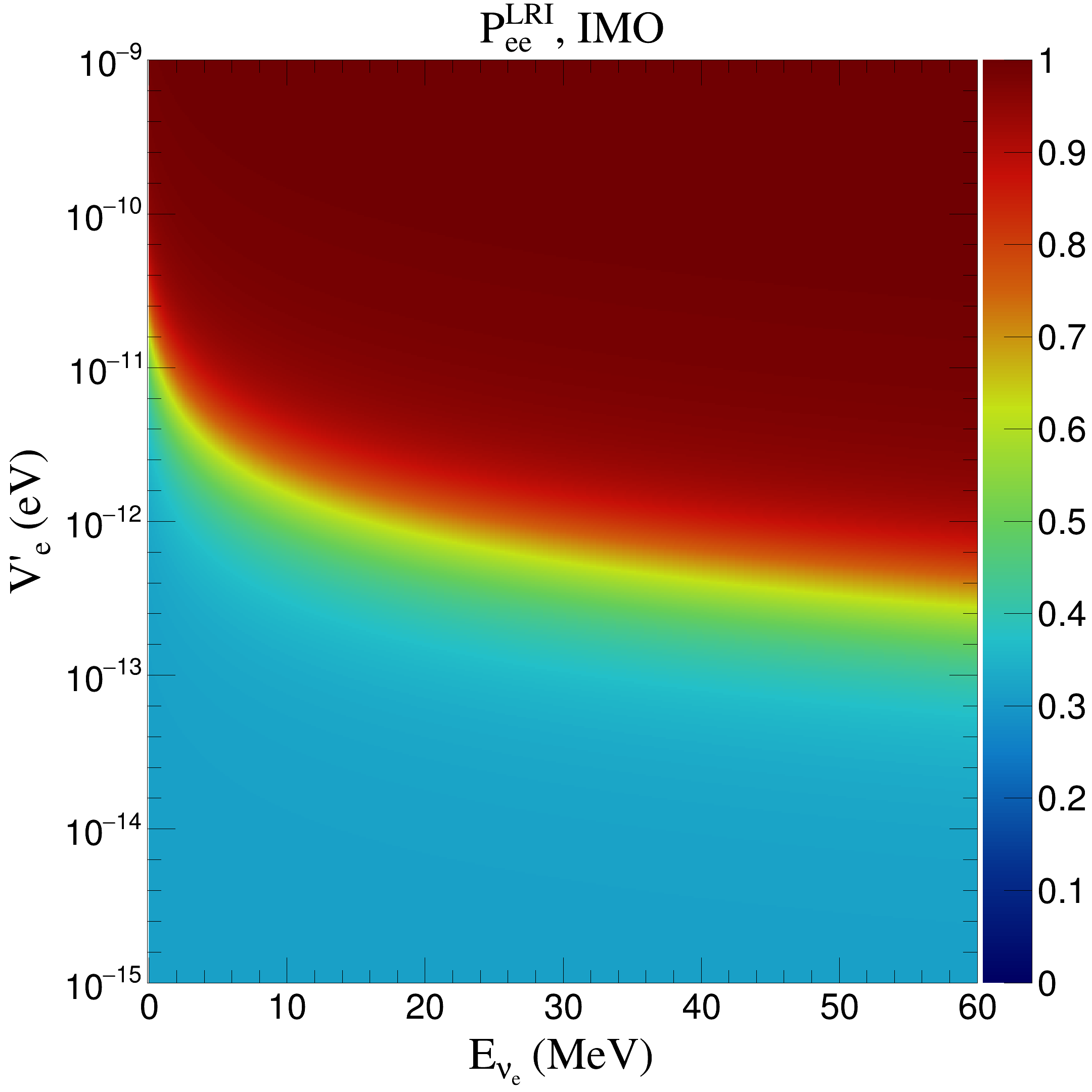}
	\mycaption{These plots show the electron-neutrino survival probability, $P^{\rm LRI}_{ee}$, indicated by the colour scale, as a function of neutrino energy (x-axis) and the effective long-range interaction potential, $V'_e$ (y-axis), evaluated at the Earth. The left (right) panel corresponds to the normal (inverted) mass-ordering scenario. }
	\label{fig:lriP}
\end{figure}

Figure~\ref{fig:lriP} demonstrates the electron-neutrino survival probability $(P^{\rm LRI}_{ee})$ at the Earth as a function of neutrino energy, as a function of the effective long-range interaction potential and the neutrino energy. The calculations employ the current global best-fit values of the oscillation parameters~\cite{Esteban:2020cvm,deSalas:2020pgw,Esteban:2024eli}.
The resulting survival-probability patterns reveal that the variation of the survival probability is larger in the NMO case (ranging from 0.022 to 1.0) than that in the IMO case (ranging from 0.3 to 1.0). Thus, the difference between the two orderings highlights the importance of precise flavour measurements during the neutronization burst. Future determination of the neutrino mass ordering will therefore enhance the interpretation of SN-neutrino searches for long-range interactions.

In summary, the existence of a background long-range potential can modify the standard flavour signatures expected from a galactic SN, and partially mimic the effects normally associated with the neutrino mass ordering. Precise measurements of the neutronization burst therefore offer a unique opportunity to probe ultraweak long-range forces and to test the robustness of standard SN neutrino flavour evolution.

\section{Sensitivity to long-range interactions with DUNE}
\label{sec:DUNEinto}
The Deep Underground Neutrino Experiment, with its 40 kt liquid-argon time projection chamber (LArTPC), is expected to provide unprecedented sensitivity to the $\nu_e$ component of a galactic core-collapse SN~\cite{Acciarri:2016crz,DUNE:2020zfm}.

The dominant detection channel is the charged-current interaction $\nu_e \,+\, ^{40}{\rm Ar}\; \to\; ^{40}{\rm K}^{*} \,+\, e^-$. The outgoing electron carries information about the incident neutrino energy, while the excited ${\rm K}^{*}$ nucleus subsequently de-excites through the emission of low-energy photons. Together, these visible final-state particles form the basis of the neutrino-energy reconstruction. The charged-current interaction threshold on argon imposes a minimum neutrino energy of $E_\nu^{\rm min}=5.25$ MeV, which is adopted throughout the analysis.
The interaction and final-state particle production are simulated using the MARLEY event generator~\cite{Gardiner2018}. Throughout this work, we assume a finite detector resolution incorporated through a Gaussian smearing procedure~\cite{Abi:2018dnh, Acciarri:2018myr},
\begin{equation}
	\sigma_E = 0.11\sqrt{E/\text{MeV}} + 0.2\,(E/\text{MeV}).
\end{equation}
The expected timing resolution of DUNE is of order ${\cal O}(10~{\rm ns})$ and is therefore negligible compared with the characteristic millisecond timescale of the neutronization burst. 

The differential event rate as a function of reconstructed energy $E^r$ and post-bounce time $t$ can be written as
\begin{equation}
	\frac{d^2 N(E^{r},t)}{dt\, dE^{r}}
	 = N_{\rm tg}\int dE^{t}\;
	f_{\nu_{\alpha}}(E^{t},t)\;
	\sigma_{\alpha}(E^{t})\;
	\mathcal{R}(E^{t},E^{r}),\\
\end{equation}
Here $N_{\rm tg}$ denotes the number of argon targets in the fiducial volume, $E^{t}$ denotes the true neutrino energy, $f_{\nu_\alpha}$ is the oscillated neutrino flux at Earth, and $\sigma_\alpha$ is the corresponding interaction cross section.
\begin{figure}[t]
	\centering
	\includegraphics[width=0.485\linewidth]{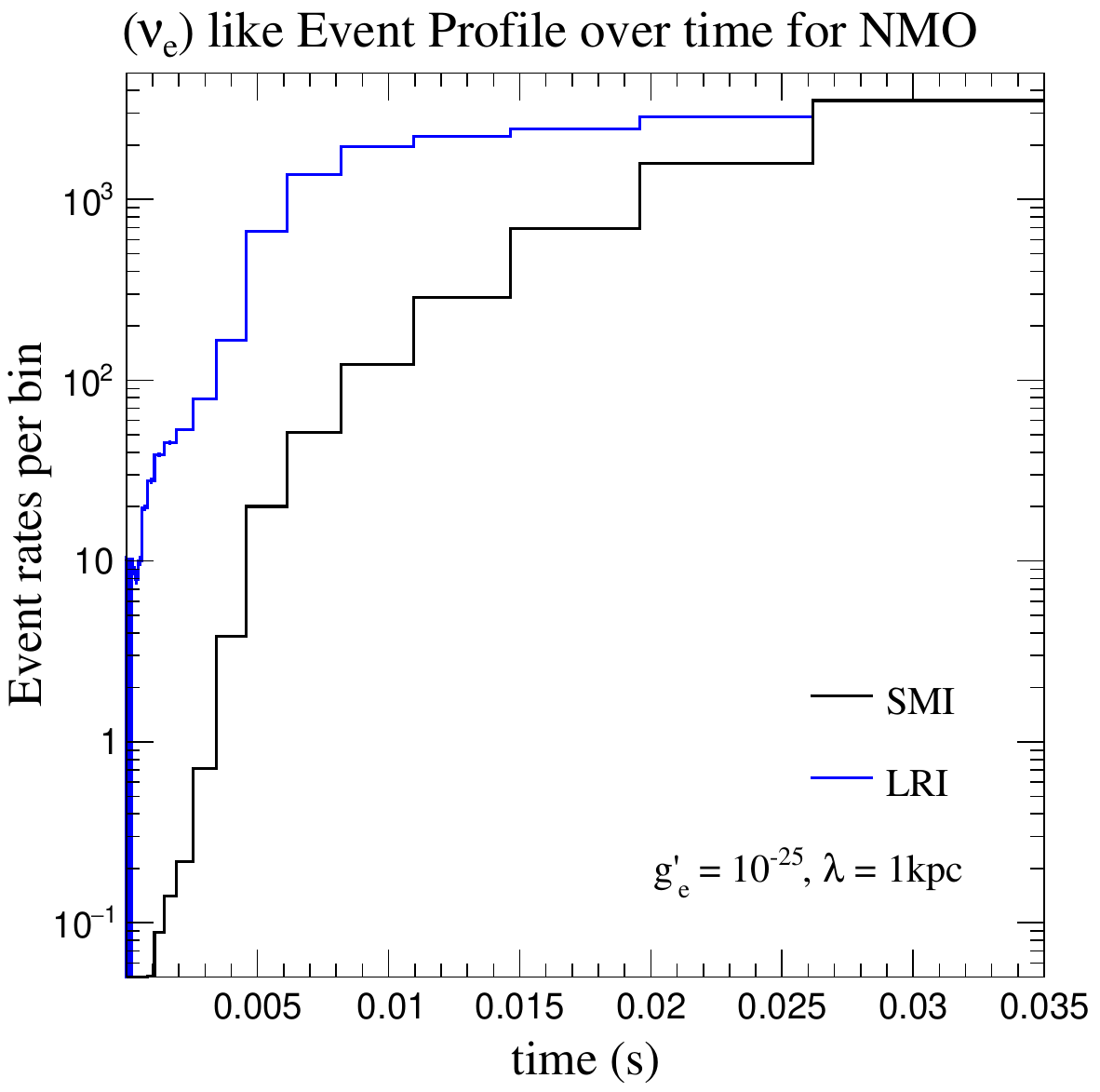}\quad
	\includegraphics[width=0.485\linewidth]{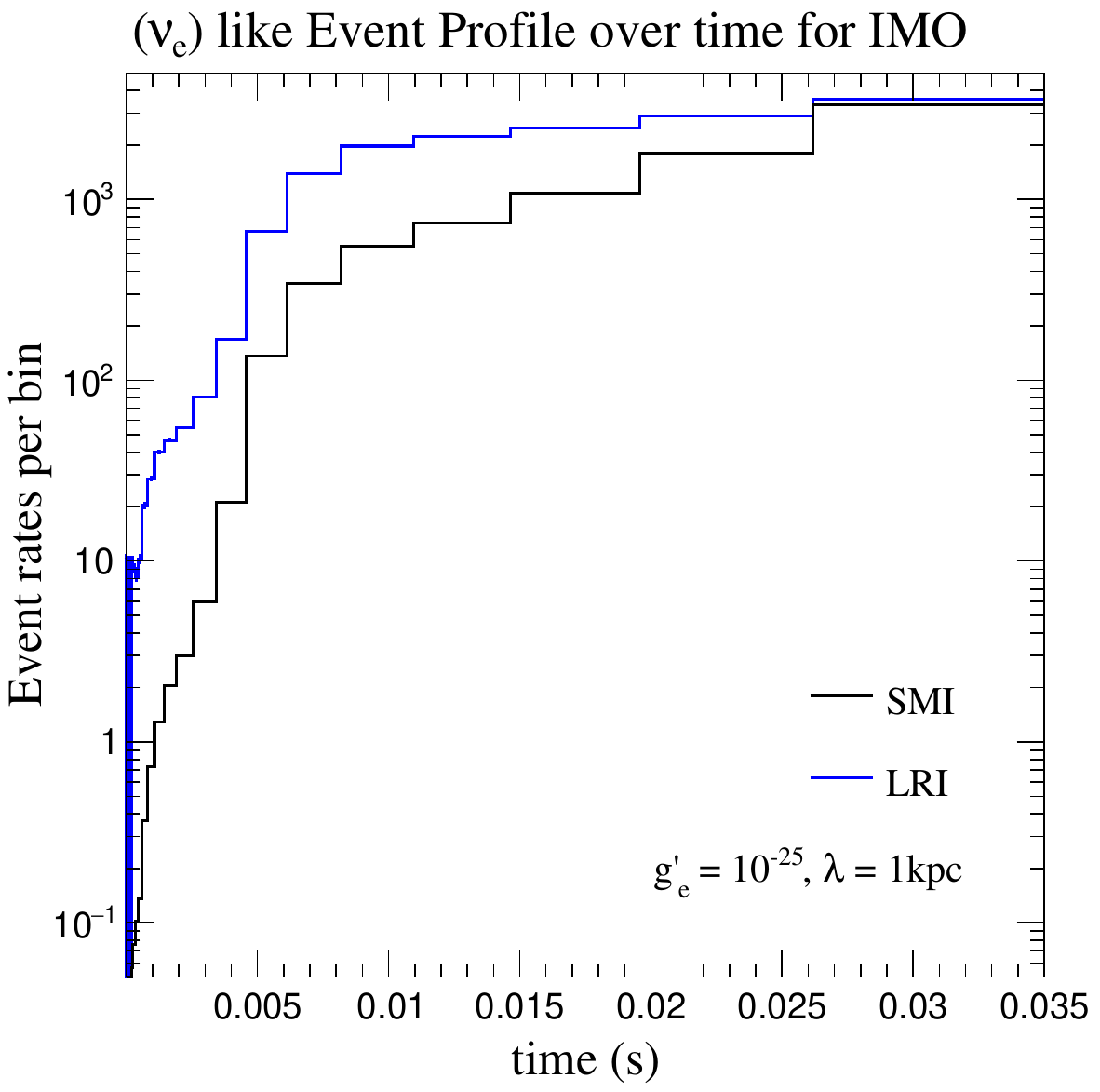}\\
	\vspace{0.215cm}
	\includegraphics[width=0.485\linewidth]{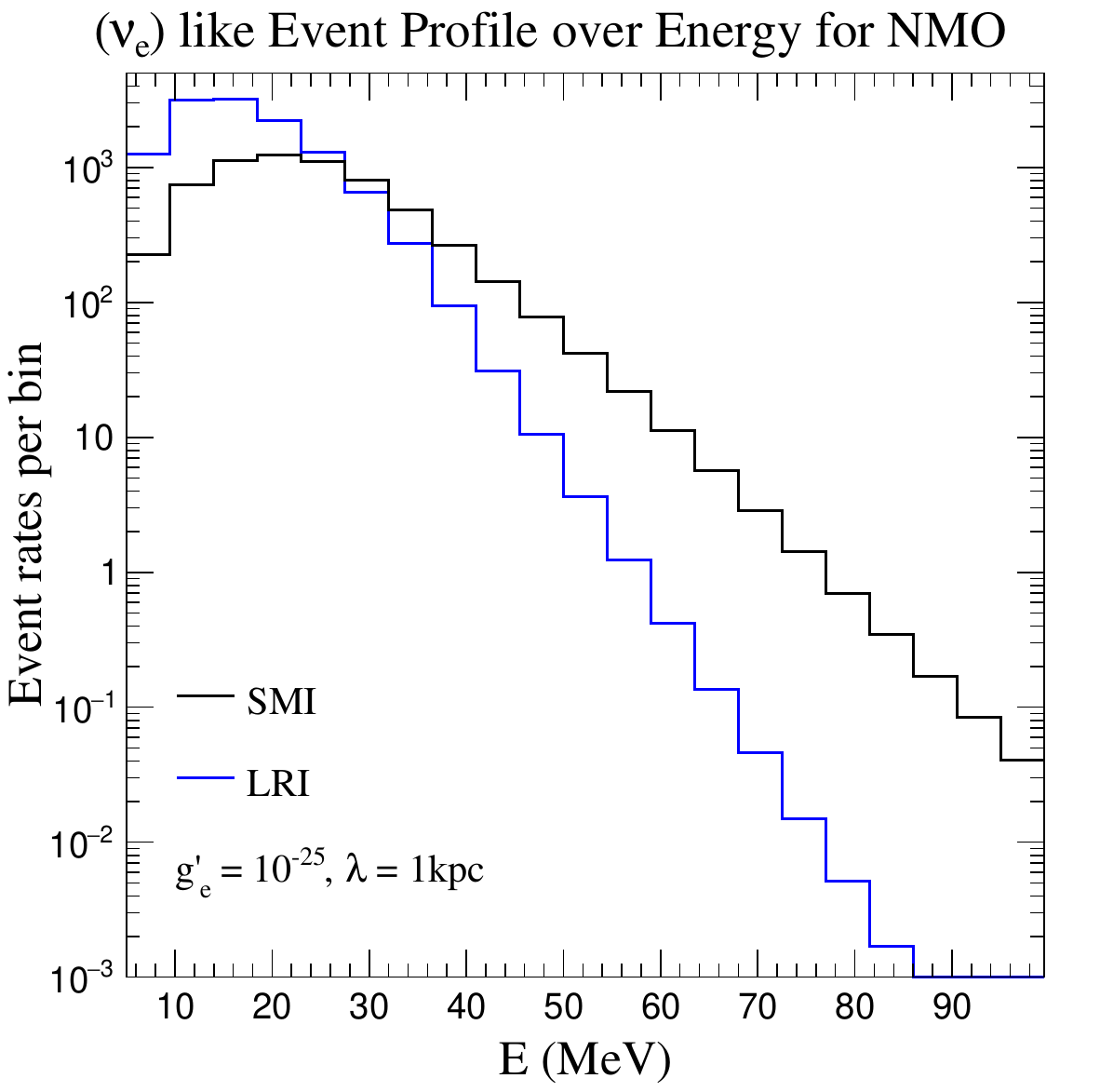}\quad
	\includegraphics[width=0.485\linewidth]{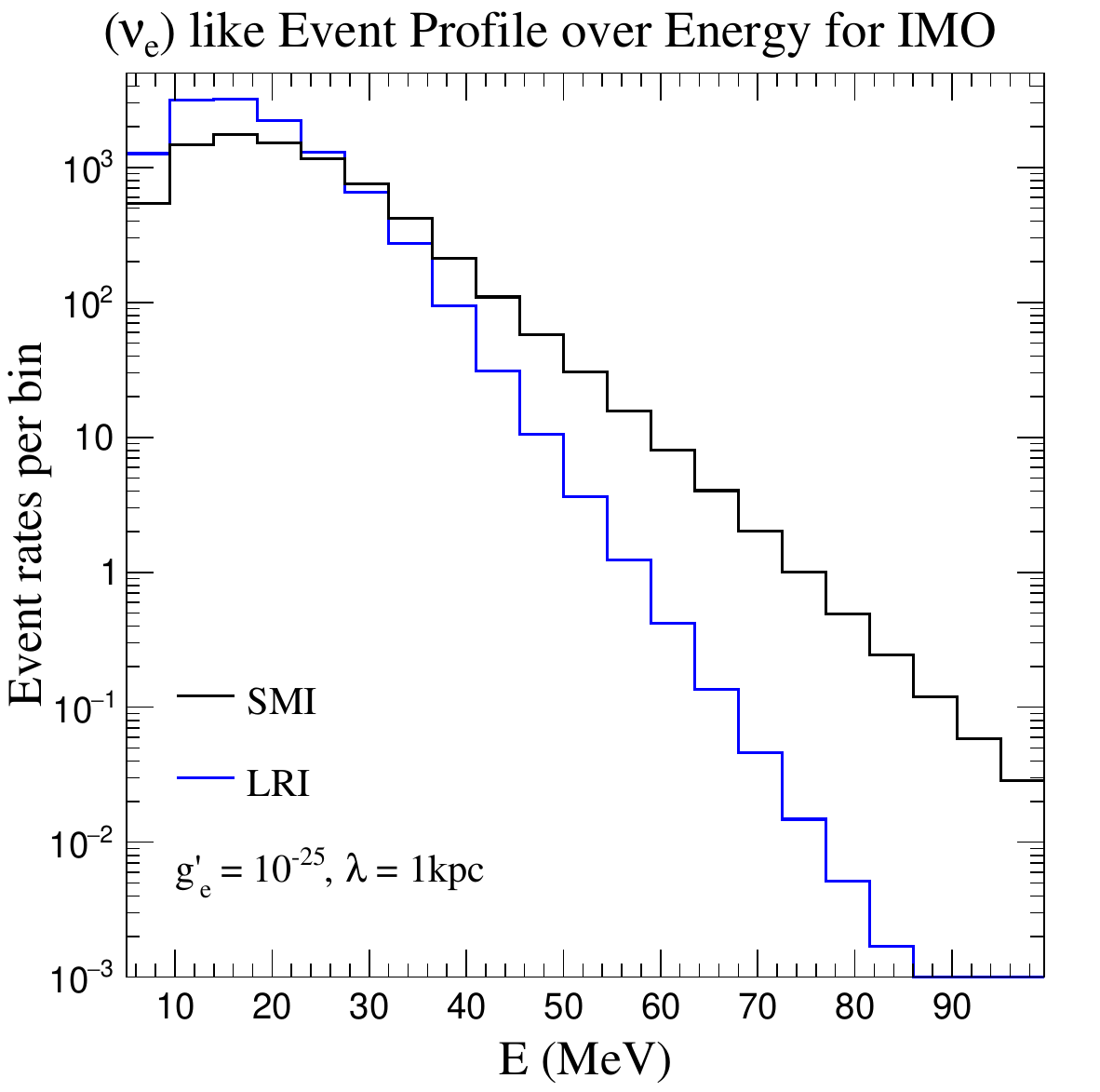}
	\mycaption{The reconstructed $\nu_e$ like neutrino event rates per bin distributions over time (upper-panel) and energy (lower-panel) spans. The source of the neutrino is taken to be Betelgeuse. The black coloured histograms illustrate the event-rate per bin profile for SMI, whereas the blue ones demonstrate the LRIs scenarios. The left and right panels consider NMO and IMO cases, respectively. }
	\label{fig:evts}
\end{figure}
The detector response function is modeled as
\begin{equation}
    \mathcal{R}(E^t,E^r) 
	 = \frac{1}{\sqrt{2\pi\sigma_E}}
	\exp\!\left[
	-\frac{\left(E^t-E^r\right)^2}
	{2\,\sigma^2_E}
	\right],
\end{equation}
which describes the migration between true and reconstructed neutrino energies.

For a nearby SN such as Betelgeuse ($\alpha$-Orion), located at a distance of approximately $168~{\rm pc}$, the expected event statistics ($N \sim 10^5$ at DUNE during the neutronization burst) are extremely large. This enables a simultaneous analysis of both temporal and spectral information contained in the neutronization burst. For such a case study, we use 1D (spherically symmetric) hydrodynamical simulation of a $25\,M_\odot$ progenitor model by the Garching group~\cite{Garching} for mimicking core-collapsed neutrinos burst from the Betelgeuse. The two panels of figure~\ref{fig:evts} show the reconstructed event rates per bin in a DUNE-like detector as a function of time as well as energy for both standard flavour evolution (black) and scenarios including long-range interactions (blue). The time axis is divided into logarithmically spaced bins covering the interval from $1~\mu{\rm s}$ to $35~{\rm ms}$, while energy is divided into eighteen uniform bins of width $4.5~{\rm MeV}$, spanning the range from $5.5\,$MeV to $100\,$MeV. 

We note that the presence of long-range interactions modifies both the overall normalization and the spectral shape of the reconstructed event distribution. For the benchmark choice $g'_e=10^{-25}$ and $\lambda=1~{\rm kpc}$, we find that the LRI potential at the Earth is dominated by the Solar contribution. The Earth and Moon provide only subleading corrections, whereas the Milky Way contribution is negligible. Thus, the total potential is controlled almost entirely by the Sun. These distortions arise from changes in the electron-neutrino survival probability discussed in the previous section. Consequently, the combination of temporal and spectral information provides a powerful handle for distinguishing long-range interaction scenarios from standard flavour evolution.
The large event statistics expected from a nearby core-collapse supernova therefore allow DUNE to operate as a precision probe of flavour-dependent long-range interactions, with sensitivity extending well beyond what can be inferred from the integrated event rate alone.

\section{Results and Discussions}
\label{sec:result}
Having established the impact of flavour-dependent long-range interactions on the reconstructed neutronization-burst signal, we now quantify the sensitivity of 40 kt LArTPC detector at DUNE to the coupling $g'_e$ for a given value of $m_{Z'}$. Our goal is to determine the extent to which deviations from standard flavour evolution can be identified using the temporal and spectral information contained in the observed event distribution.

The analysis is performed in the two-dimensional space of reconstructed neutrino energy and post-bounce time described in the previous section. We simulate true $\nu_e$-like events $(N_{ij}^{\rm true})$using only SMI for both NMO and IMO. For testing the LRI scenario, we reconstruct the time and energy distributions of the $\nu_e$-like events $(N_{ij}^{\rm test})$ by varying the couplings $(g'_e)$ for different values of $\lambda$ (or equivalently, $m_{Z'}$). Here, the net LRI potential is induced by the total electron abundance in the SN, the Milky Way galaxy, the Sun, and the Earth\footnote{In this section, we focus on the range $\lambda \sim 10^{-10}$ kpc (where the LRI potential due to the Earth dominates over all other contributions) to $\lambda \sim 10^{-1}$ kpc (distance of Betelgeuse from the Earth). In this region, the combined contribution of the Milky Way as well as the extragalactic sources may be neglected. The results for a wider range of $\lambda$ will be presented in the next section.}.
We adopt a frequentist approach to get the median sensitivity of the detector, assuming a Gaussian $\chi^2$ distribution, minimized over systematic uncertainties:
 \begin{align}
	&\chi^2(g'_e) 
	= \min_{\xi_k}\left[\;\;\sum_{i=1}^{N_t} \sum_{j=1}^{N_E}
	\bigg(\frac{N_{ij}^{\rm test}(g'_e, \xi_k)-N_{ij}^{\rm true}}{\sigma_{ij}}\bigg)^2\,+\, \sum_{k=1}^{3}\xi^2_k\;\right]\,,
\end{align} 
where the binning indices $i$ and $j$ label the time and reconstructed energy bins, respectively, and $\sigma_{ij}^2 = N_{ij}^{\rm true}$. The reconstructed number of events in a bin have been written as $N_{ij}^{\rm test} = \bigg(1+\sum_{k=1}^{3}\pi^k_{ij}\,\xi^2_k\bigg)\,N_{ij}^{(0)\,\rm test}$, where the variable $N_{ij}^{(0)\,\rm test}$ stands for the predicted number of reconstructed events without any systematic uncertainties. However, possible systematic uncertainties ($\pi^k_{ij}$) can shift the median of the pure predicted events. Therefore, we employ the pull-variable approach to handle such systematic fluctuations~\cite{fogli:2002pt}. Here, we consider a linearized parameterization of three systematic uncertainties, viz. 10\% uncertainty on flux normalization\footnote{Since the mass and distance of Betelgeuse is well-known, the model dependence of the time and energy distributions of neutrino emission during the neutronization burst is small~\cite{Kachelriess:2004ds}. We have also checked that our results do not change appreciably if the flux normalization uncertainty is taken to be 20\% instead.}, 10\% uncertainty on cross-section and an overall 5\% systematic uncertainty, which we quantify in terms of a set of pull variables ($\xi_k$). The minimization over the pull parameters ensures that the analysis remains insensitive to small overall normalization shifts and instead focuses on distortions of the event distribution induced by long-range interactions. 

The sensitivity of DUNE to the LRI scenario can be computed through: 
\begin{align}
	\Delta \chi^2 (g'_e) & = \chi^2 (g'_e) - \chi^2_0,
\end{align}
where $\chi^2_0$ represents the standard statistical fluctuations associated with the current detector setup. Since the number of events is overwhelmingly high, we can assume this as Asimov dataset~\cite{cowan:2010js}, and the value of $\chi^2_0$ can be suppressed to null.

In this study, we take the three-neutrino oscillation parameters ($\theta_{12} = 34.5^\circ$, $\theta_{23} = 45^\circ$, $\theta_{13} = 8.5^\circ$, $\Delta m^2_{21} = 7.5\times10^{-5}$ eV$^2$ and $\Delta m^2_{31} = 2.5\times10^{-3}$ eV$^2$), consistent with the fit values given in NuFit 6.0~\cite{Esteban:2024eli}. We keep these parameters fixed throughout our analysis since there is no direct impact of the current mixing parameter uncertainties associated with $\theta_{23}$ and $\delta_{\rm CP}$ (which have relatively large uncertainties) on the relevant probabilities. However, we perform our analysis for both the possibilities of mass ordering (NMO and IMO).

\begin{figure}[t]
	\centering
	\includegraphics[width=0.5\linewidth]{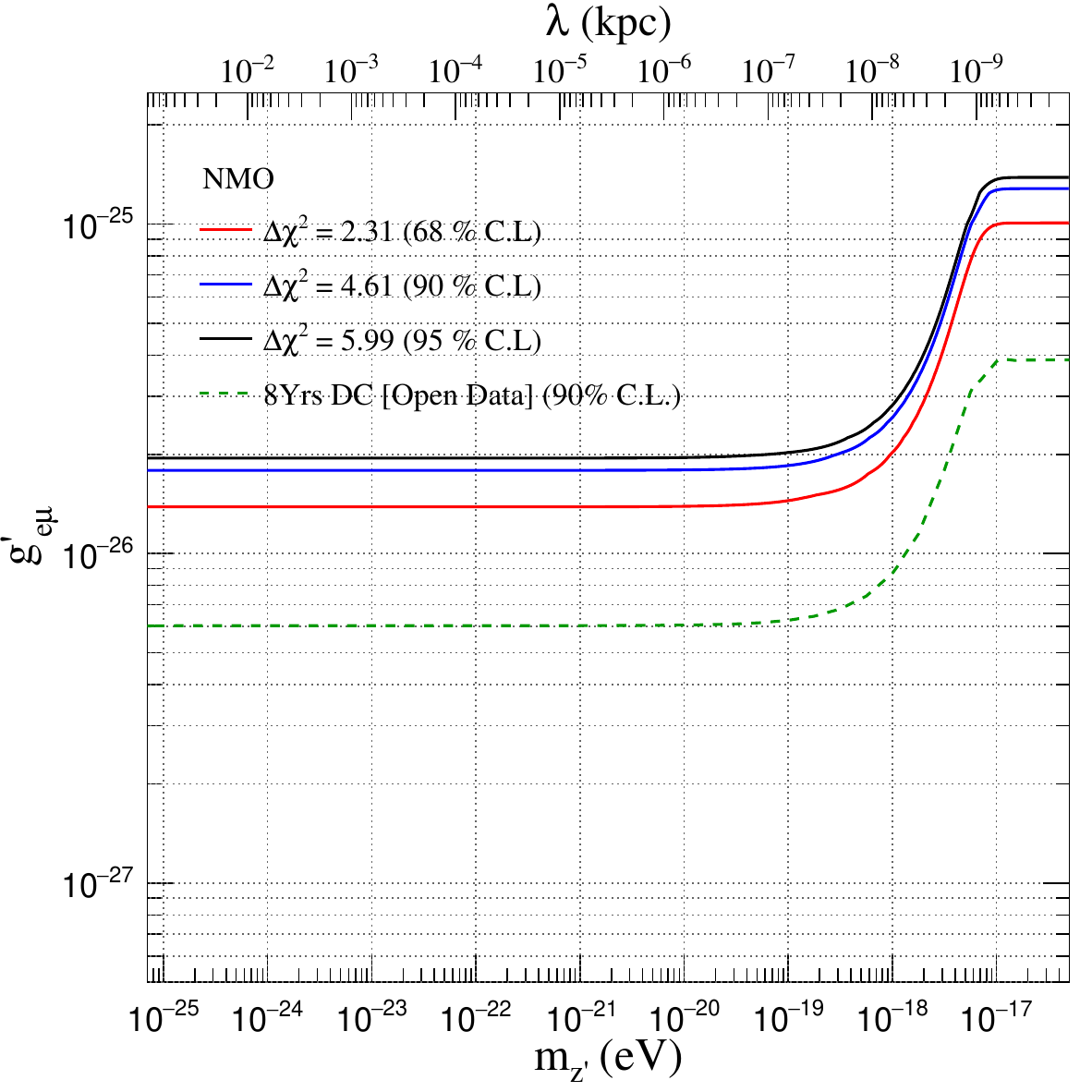}~
	\includegraphics[width=0.5\linewidth]{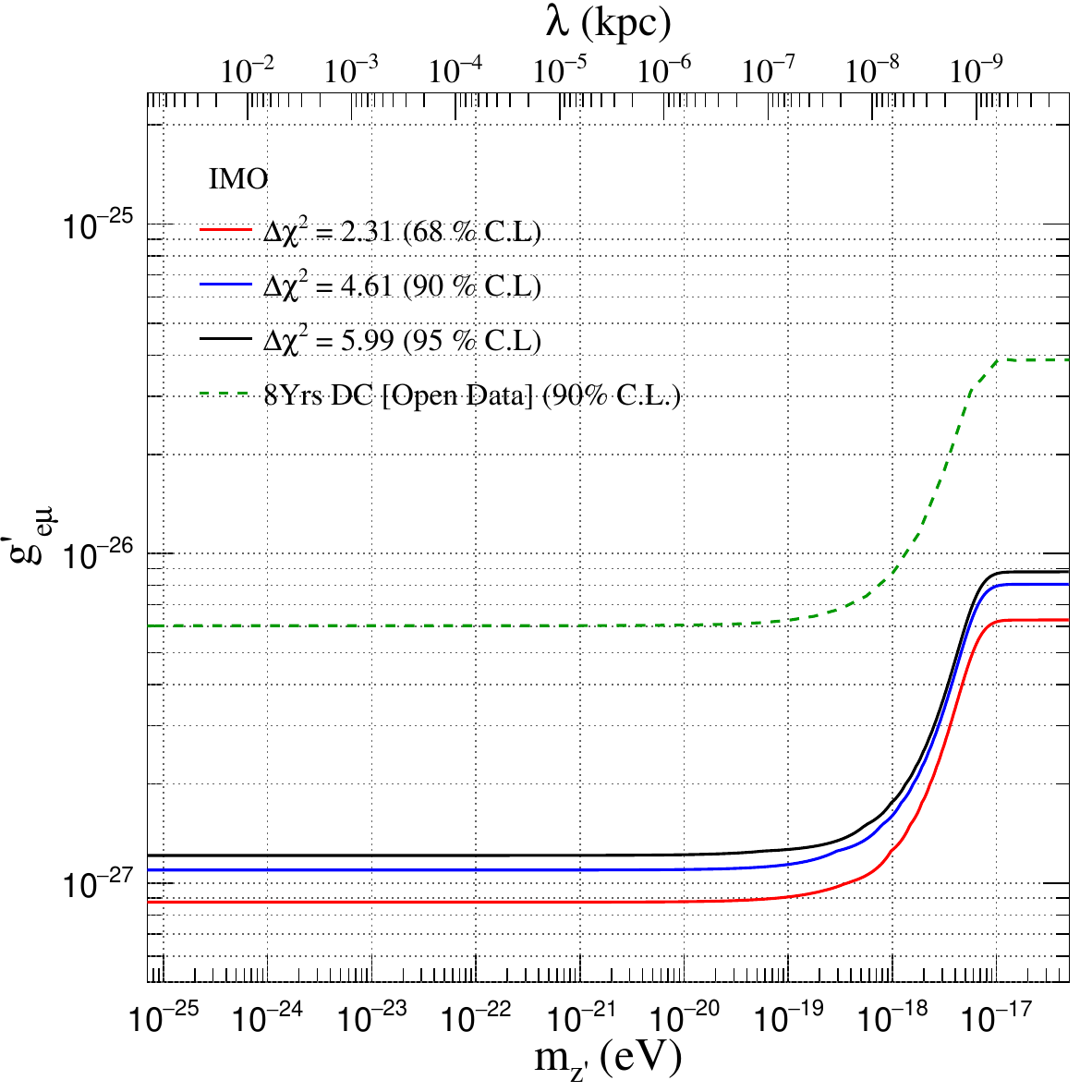}
	\mycaption{The sensitivity of 40 kt LArTPC detector at DUNE in the plane of $g'_e$ and $m_{z'}$, if the neutronization burst from Betelgeuse is observed . The left (right) panels consider the NMO (IMO) as mass-ordering in both true and test hypotheses. The contours are drawn with 68\% C.L. (red), 90\% C.L. (blue) and 95\% C.L. (black) of the $\Delta\chi^2$ values for one-parameter estimation. The green dashed curve represents the constraint imposed on the parameter space, using multi-GeV atmospheric neutrinos with an 8 years of Open-Data base registered at IceCube (DeepCore), see reference~\cite{Garg:2026gwx}.}
	\label{fig:rslt1}
\end{figure}

\noindent
Figure~\ref{fig:rslt1} shows the sensitivity of DUNE to neutrino LRI using the neutronization-burst signal from a Betelgeuse-like SN candidate occuring at $168~$pc. The red, blue and black contours correspond to $\Delta\chi^2$ values $1.00$ ($68\%$ C.L.), $2.71$ ($90\%$ C.L.) and $3.84$ ($95\%$ C.L.), respectively, for one-parameter estimation. The sensitivity depends strongly on the interaction range. For very short interaction lengths, $\lambda\ll L$, the Yukawa suppression factor strongly reduces the long-range potential, therefore, rendering interaction effectively invisible. In the opposite regime, the potential approaches its long-range limit and the sensitivity becomes controlled primarily by the coupling strength.

The transition between these regimes gives rise to the characteristic shape of the exclusion contours. As the interaction range becomes comparable to the relevant astrophysical distance scale, additional electron reservoirs begin to contribute to the effective potential, producing a rapid increase in sensitivity. Once the interaction length exceeds the size of the dominant source population, the potential saturates and further increases in $\lambda$ no longer lead to substantial improvements.

A notable feature of the results is the difference of almost an order of magnitude in the sensitivities between the NMO (left) and IMO (right). This behaviour originates from the distinct flavour composition of the neutronization burst after adiabatic propagation through the SN envelope. As discussed before, for the NMO, the emitted $\nu_e$ flux exits the star predominantly as the vacuum mass eigenstate $\nu_3$. Hence, as long as there is no signature of LRI, $P_{ee}$ would be approximately 0.022. Similarly, for IMO, the $\nu_e$ arrives predominantly as $\nu_2$, and hence $P_{ee}$ would be approximately 0.30. Thus, statistics is expected to dominate in the IMO and give a stronger sensitivity than that in NMO. We further find that for the benchmark case considered, the expected sensitivity in IMO can exceed the current constraint imposed using multi-GeV atmospheric neutrino data in IceCube (DeepCore)~\cite{Garg:2026gwx}. However, if the mass ordering turns out to be normal, we expect the current constraints from IceCube to be stronger.

Finally, we want to highlight that a Betelgeuse-like SN candidate at around 0.1 kpc allows us to probe such extreme values of ($g'_e - m_{Z'}$) due to the enhanced statistics as compared to a more distant SN. Nevertheless, even for a more distant galactic SN, the neutronization burst remains sufficiently clean that meaningful constraints on long-range interactions can still be obtained.

\section{Concluding remarks}
\label{sec:cnrmk}
\begin{figure}[ht]
	\centering
	\includegraphics[width=0.775\linewidth]{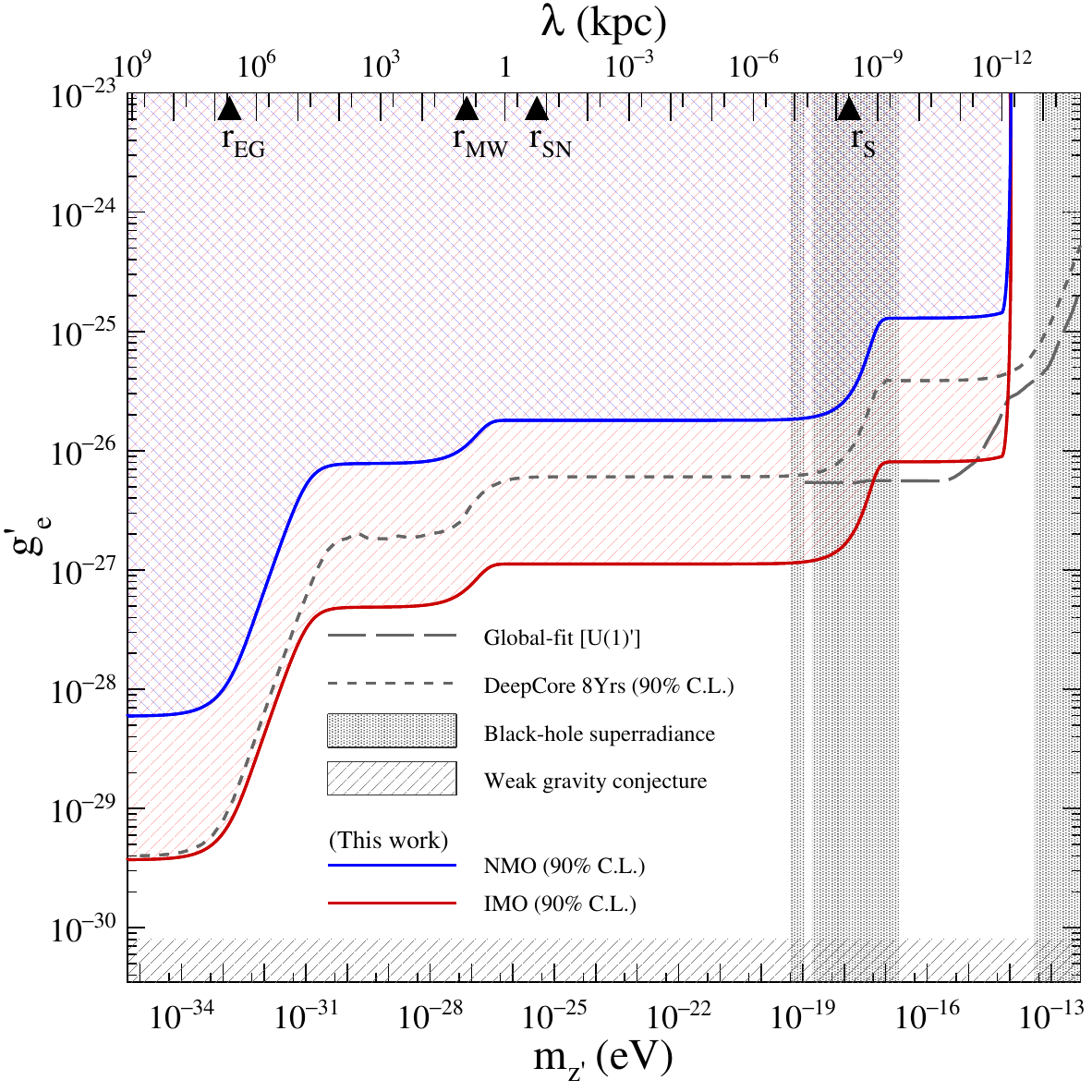}
	\mycaption{Global comparison of existing constraints and projected sensitivities in the $(g'_e,\,m_{Z'})$ parameter space if the neutronization burst from Betelgeuse is observed. The quantities $r_{\rm S}$, $r_{\rm SN}$, $r_{\rm MW}$ and $r_{\rm EG}$ denote the characteristic order-of-magnitude radial distances from the Earth to the Sun, Betelgeuse, Milky Way and extragalatic regions, respectively. The blue (NMO) and red (IMO) shaded regions show the projected sensitivity for an anomaly-free $L_e-L_\mu$ LRI scenario, obtained in this work. The dotted contour shows the bound derived from eight years of multi-GeV atmospheric-neutrino data collected by IceCube-DeepCore~\cite{Garg:2026gwx}, while the wide-dashed contour corresponds to constraints from a global neutrino-oscillation analysis within a minimal $U(1)'$ extension of the Standard Model~\cite{Coloma:2020gfv}. The gray vertical hatched region indicates the parameter space excluded by black-hole superradiance~\cite{Baryakhtar:2017ngi}. The gray horizontal shaded band represents the upper bound on the gauge coupling inferred from quantum-gravity considerations, based on the requirement that gravity remains the weakest force~\cite{Arkani-Hamed:2006emk}.}
	\label{fig:GlbLt}
\end{figure}
Flavour-dependent long-range leptonic interactions arise naturally in a wide class of extensions of the Standard Model, containing ultra-light gauge bosons associated with additional symmetries. 
Due to the small mass of the mediators, the large interaction range can be astronomical and can generate observable effects in neutrino flux from astrophysical sources. In this regard, a core-collapse supernova provides a perfect target to probe this new physics with MeV energy neutrinos. The naturally ultra-long baselines provide us with a unique opportunity to probe extremely small coupling regimes, which are otherwise inaccessible to conventional laboratory experiments.

The neutronization burst provides an exceptionally clean environment for probing new physics in the neutrino sector. Due to the dominance of electron capture during the early post-bounce phase, the emitted neutrino flux consists almost entirely of electron neutrinos. The contribution of collective flavour conversions is negligible during the burst. Furthermore, adiabatic propagation ensures that the electron neutrinos propagate as the heaviest matter/mass eigenstate enroute to the Earth. As a consequence, the observable signal becomes highly sensitive to modifications of the effective propagation Hamiltonian induced by long-range interactions.

In this work, we have investigated the sensitivity of neutronization-burst neutrinos to long-range interactions arising out of an underlying $U(1)'_{L_e-L_\mu}$ symmetry. As a benchmark, we assumed the progenitor to be a Betelgeuse-like star, located at a distance of 168 pc. We showed that long-range potential due to solar electrons dominate the neutrino flavour propagation and can suppress flavour mixing altogether. This can modify the effective mixing angles governing the electron-neutrino survival probability and thereby enhance the neutrino signal at the Earth.

To analyze the impact of these long-range interactions, we calculated the reconstructed neutronization-burst signal in both energy and time using 40 kt LArTPC detector at the Deep Underground Neutrino Experiment. The resulting event distributions were analysed using a binned statistical framework that incorporates detector resolution and systematic uncertainties. We found that for a Betelgeuse-like SN going off at a distance of around 0.1 kpc, DUNE can provide a valuable probe of flavour-dependent long-range forces and yield constraints complementary to those obtained from laboratory experiments, oscillation measurements, and astrophysical observations. Clearly, the enhanced event statistics associated with a nearby source substantially improves the sensitivity to long-range interactions.

We summarize our projected sensitivities together with the existing constraints in the $(g'_e,m_{Z'})$ parameter space in figure~\ref{fig:GlbLt}. Here, we extend the explored parameter space to a broader range of mediator masses by explicitly incorporating the contribution of the extragalactic electron population to the LRI potential at the Earth. The corresponding extragalactic potential, $\mathcal{V}_{\rm EG}$, is computed by integrating the redshift-dependent cosmic star formation rate density~\cite{Horiuchi:2008jz}. Our results demonstrate that the sensitivity depends strongly on both the interaction range and the neutrino mass ordering. For interaction ranges smaller than the Earth's radius, the long-range interaction potential remains subdominant to the standard oscillation Hamiltonian, leading to a rapid loss of sensitivity\footnote{The Earth-matter effects depend on the arrival direction of SN neutrinos and their propagation path through the Earth; it is expected to modify the oscillation probabilities by less than $5\%$ for neutrino energies of $\mathcal{O}(10\,\mathrm{MeV})$.}. As the interaction range increases, progressively larger astrophysical electron reservoirs contribute to the effective potential, resulting in a corresponding improvement in the sensitivity before eventually reaching the asymptotic long-range regime. This gives rise to the characteristic step-like features, reflecting the successive contributions from the Earth, the Sun, the SN progenitor, the Milky Way, and finally the extragalactic electron population. The projected sensitivity also exhibits a dependence on the neutrino mass ordering, arising from the different mass-eigenstate composition of the neutronization-burst neutrino flux after adiabatic flavour evolution inside the SN. In particular, for the inverted mass ordering, DUNE can surpass the current constraints derived from eight years of IceCube (DeepCore) data over most of the parameter space.

Thus, the neutronization-burst neutrinos constitute a powerful and theoretically robust probe of ultra-light leptophilic interactions. Unlike many other astrophysical observables, the initial flavour composition of the neutronization burst is both well understood and only weakly dependent on uncertainties in the SN explosion mechanism. The future observation of a galactic SN by DUNE, therefore, has the potential to probe flavour-dependent long-range interactions with unprecedented precision and to provide a unique window into new forces acting in the lepton sector.

While the sensitivity obtained in this work relies on the occurrence of a nearby supernova such as Betelgeuse, the underlying methodology is considerably more general. Future improvements in detector technologies, low-energy event reconstruction, and machine-learning-based analysis techniques are expected to enhance the sensitivity to supernova neutrinos across a broad range of observables. In the longer term, precision measurements of the diffuse supernova neutrino background may provide an additional avenue for exploring flavour-dependent interactions over cosmological baselines, offering a complementary probe of new physics beyond the Standard Model.

\section*{Acknowledgements}
\label {subsec:ackw}
We acknowledge the XVIII Workshop on High Energy Physics Phenomenology\\ (WHEPP-2025) for providing a platform for discussions during the initial stages of the work. We also thank the High-Performance Computing facilities (NSM-Param Rudra) at IIT Bombay which has been used to perform the numerical simulations. We further thank Sudipta Das and Pragyanprasu Swain for useful discussions. AD acknowledges support by the Department of Atomic Energy, Government of India, under Project Identification Number RTI 4012, and funding from the J.C. Bose Grant ANRF/JBG/2025/000265/PS of the Anusandhan National Research Foundation (ANRF), Government of India. SS thanks IIT Bombay for funding to this project through IPDF fellowship. MS acknowledges support from the Early Career Research Grant by Anusandhan National Research Foundation (project number ANRF/ECRG/2024/000522/PMS).

\bibliographystyle{JHEP}
\bibliography{LRI_SN}

\end{document}